\documentclass[12pt,a4paper,final]{iopart}
\usepackage{iopams}  
\usepackage{graphicx}
\usepackage[breaklinks=true,colorlinks=true,linkcolor=blue,urlcolor=blue,citecolor=blue]{hyperref}

\usepackage{natbib}

\newcommand{\av}[1]{\langle{#1}\rangle}

\begin{document}

\title{Relativistic numerical cosmology with Silent Universes}

\author{Krzysztof Bolejko}
\address{Sydney Institute for Astronomy, School of Physics, A28, The University of Sydney, NSW, 2006, Australia}
\ead{krzysztof.bolejko@sydney.edu.au}

\begin{abstract}
Relativistic numerical cosmology is most often based either on the exact solutions of the Einstein equations, or perturbation theory, or weak-field limit, or the BSSN formalism. The Silent Universe provides an alternative approach to investigate relativistic evolution of cosmological systems. 
The silent universe is based on the solution of the Einstein equations in 1+3 comoving coordinates with additional constraints imposed. These constraints include: the gravitational field is sourced by dust and cosmological constant only, both rotation and magnetic part of the Weyl tensor vanish, and the shear is diagnosable. This paper describes 
the code \textsl{simsilun} (free software distributed under the terms of the reposi
 General Public License),
which implements the equations of the Silent Universe. {  The paper also discusses applications of the Silent Universe and
it uses the Millennium simulation to set up the initial conditions for the code \textsl{simsilun}. The simulation obtained this way consists of 16,777,216 worldlines, which are evolved from $z=80$ to $z = 0$.
Initially, the mean evolution (averaged over the whole domain) follows the evolution of the background $\Lambda$CDM model. However, once the evolution of cosmic structures becomes nonlinear, the spatial curvature
evolves from $\Omega_K =0$ to $\Omega_K \approx 0.1$ at the present day. 
The emergence of the spatial curvature is associated with $\Omega_M$ and $\Omega_\Lambda$ 
being smaller by approximately $0.05$ compared to the $\Lambda$CDM. }
\end{abstract}

\vspace{2pc}
\noindent{\it Keywords}: cosmology: theory -- cosmological parameters -- large-scale structure of Universe -- methods: numerical

%\maketitle

\section{Introduction}\label{intro}

The year 2017 marked the 100th anniversary of relativistic cosmology that started with Einstein's cosmological model \citep{1917SPAW.......142E} and de Sitter's lecture {\em On the relativity of inertia} delivered on 31th of March 1917, at the Meeting of the Royal Netherlands Academy of Arts and Sciences \citep{1917KNAB...19.1217D}. The following years brought new cosmological solutions of the Einstein equations, such as the solutions found by \citet{1922ZPhy...10..377F,1924ZPhy...21..326F} and \citet{1927ASSB...47...49L}.
These solutions were based on the assumption of homogeneity and isotropy, 
further investigated by \citet{1929PNAS...15..822R,1933RvMP....5...62R} and
 \citet{Walker1935} (hence the name {\em FLRW models}). 
The FLRW models also include, as a spatial case, the spatially flat Einstein-de Sitter model \citep{1932PNAS...18..213E}.

Inhomogeneous but isotropic solutions of the Einstein equations were first found by  \citet{Lemaitre}, and then further studied by \citet{Tolman} and  \citet{1947MNRAS.107..410B} (hence the name {\em LTB models}).
These spherically symmetric models were then generalised by \citet{1967JMP.....8.1171E}  to include other $G_3/S_2$-symmetric spacetimes.  Soon  \citet{1969CMaPh..12..108E} investigated anisotropic but homogeneous solutions of the Einstein equations, the so called {\em Bianchi models}. The most general (currently known) cosmological solution of the Einstein equations that is both inhomogeneous and anisotropic was found by  \citet{Szekeres:1975ct} and then generalised (to include gradient-free pressure) by   \citet{Szafron:1977zza}. The Szekeres-Szafron solutions are quite general and contain in special limits the LTB and FLRW models and also the Schwarzschild solution \citep{2006igrc.book.....P}.
(For a review on different inhomogeneous cosmological solutions and their applications see the monograph by \citet{1997icm..book.....K} and the review article by \citet{Bolejko:2011jc}).

In parallel to exact solutions, a perturbative approach had also been developed during this time, especially throughout 1980s, with the focus on a covariant approach to perturbations \citep{1980PhRvD..22.1882B,1989PhRvD..40.1804E}.
However, the perturbative approach
requires a stable background, and so it may not be accurate if the 
growth of nonlinear structures affects the evolution of the background, the so called cosmological {\em backreaction}  \citep{2009PhRvD..80h3525C}.
Although studies of backreaction within the perturbative schemes are being developed \citep{2012PhRvD..86b3520B,2013PhRvD..87l3503B,2015PhRvD..92b3512A}, currently most studies and applications of the perturbative schemes focus on  statistical properties of matter  on very large scales, in particular on the  nonlinear evolution of the matter power spectrum \citep{Baumann:2010tm,2015JCAP...05..007B,2015PhRvD..92l3007B,2016JCAP...07..052B}. 

In 1960s and 1970s an alternative approach to study relativistic and nonlinear cosmological systems started being developed. This approach is based on 1+3 split \citep{1971grc..conf..104E,2009GReGr..41..581E} and it proved to be quite a powerful tool to study properties of relativistic cosmological systems. The two most known developments that followed from this approach are: models with the {\em Local Rotational Symmetry} (LRS) \citep{1996CQGra..13.1099V} and {\em Silent Universes} \citep{1995ApJ...445..958B,1997CQGra..14.1151V}. The class II of the LRS models\footnote{The LRS models can be divided into 3 categories, with class I generalising the G\"odle solution and class III generalising the Bianchi models.} and silent universes have a common subset in the LTB models. The class II reduces to the LTB family when pressure and geodesic acceleration vanish ($p = 0 = \dot{u}$), and the silent universe reduces to the LTB solution when the isotropy is imposed.

After the initial period of interest around the silent universes, the enthusiasm faded. 
In mid 1990s it was hoped that this approach would allow to trace the evolution of cosmological systems far into the nonlinear regime.
However, limitations due to diagonalisable shear tensor and vanishing magnetic part of the Weyl tensor \citep{1997PhRvD..55.5219M}, as well as the absence of rotation, showed that one cannot trace the evolution too far into the nonlinear regime \citep{2002PhRvD..66l4015E}. 
Also, at that time Newtonian numerical simulations started gaining momentum 
proving they can be very useful to study highly nonlinear stages of evolution of cosmic structures and their properties \citep{1996ApJ...462..563N}. By mid 2000s the Newtonian simulations
provided community with synthetic universes such as for example the Millennium simulation \citep{2005Natur.435..629S}. However, $N$-body simulations are based on Newtonian cosmology and use periodic boundary conditions, which means that the global evolution must follow the FLRW evolution \citep{1997A&A...320....1B}. This issue is important for relativistic cosmologists, who since mid 1980s have been investigating the phenomenon of backreaction (for a review on backreaction and the survey of opinions see a review article by \citet{2017IJMPD..2630011B}).
The phenomenon of backreaction describes how the global evolution of the Universe is affected by the nonlinear growth of cosmic structures. So although Newtonian simulations are very successful in explaining small scale processes such as for example properties of galaxies \citep{2014Natur.509..177V,2015MNRAS.446.2744L}, they have limited applicability to studies of the global evolution of the universe that undergoes a nonlinear growth of structures \citep{2017arXiv170400703B}.

The extensions of the Newtonian $N$-body simulations to include post-Newtonian corrections have been investigated and it was shown that the mean evolution is well approximated by the Friedmannian evolution \citep{2013PhRvD..88j3527A,2014CQGra..31w4006A,2016JCAP...07..053A}. This is related to the fact that periodic boundary conditions impose a constraint on the global spatial curvature and force it to vanish \citep{2017arXiv170609309A}.
This is an important observations, because it is the global spatial curvature that has been identified as a key element of backreaction \citep{2005PhLA..347...38E,2008CQGra..25s5001B,2009GReGr..41.2017B,2011CQGra..28p5004R,2017JCAP...06..025B,2017arXiv170701800B}.

{  
Recently, \citet{2017MNRAS.469L...1R} suggested to include a Buchert-type backreaction 
to evaluate expansion rate of $N$-body domains based on their cosmological environment.
This has been extended by \citet{2017arXiv170606179R} to include the 
Relativistic Zeldovich Approximation (RZA) to describe the expansion rate of various domains within  $N$-body simulations and to make the treatment of
virialisation explicit rather than implicit.
 The RZA is a powerful tool \citep{1995PhRvD..52.5605K,1998PhRvD..57.6094M,2012JCAP...06..021R,2013PhRvD..87l3503B,2015PhRvD..92b3512A}, 
which is a general-relativistic approximation that goes far beyond standard
perturbation theory, and for example successfully describe collapsing structures and predicts the mass function that to the  first order is comparable to $N$-body
simulations, but is relativistic in origin \citep{2016arXiv160200302O,2017AcPPS..10..407O}. 

This new approach of extending the $N$-body simulations has recently sparked a debate \citep{2017MNRAS.469..744K,2017arXiv170400703B,2017arXiv170606179R}.
There are many issues that need further examination, such as
the constraints that follow from periodic boundary conditions,
the implementation of expansion rates that vary spatially among
local domains, the emergence of the average spatial curvature,
and virialisation.
Therefore, it seems that in order to thoroughly understand the backreaction phenomenon we will require a fully relativistic description of the evolution of  cosmological systems, which most likely will only be achieved by the means of numerical relativity. 
}

In the recent years, there have been serious developments of relativistic numerical cosmology  based on the Einstein toolkit \citep{2012CQGra..29k5001L} and the  Baumgarte-Shapiro-Shibata-Nakamura  (BSSN) formalism \citep{1995PhRvD..52.5428S, 1999PhRvD..59b4007B}. The results obtained by \citet{Bentivegna:2015flc}, \citet{Mertens:2015ttp}, and \citet{2017PhRvD..95f4028M}
 are impressive and encouraging. However, due to presence of shell crossings 
and due to the fact that these simulations are significantly CPU expensive, their applications is currently quite limited. 
On the  other hand, silent cosmology can provide a viable alternative, at least until the relativistic numerical cosmology is fully developed. The code \textsl{simsilun} described in Sec. \ref{code} is very fast --- the evolution of 10,000 worldliness with non-extreme densities, over $10$ Gyr, takes 1 second on a 8-core CPU machine with the openMP parallelisation. The silent universes can provide a benchmark for full numerical cosmology, at least in the mildly nonlinear regime.

{ 
The structure of the paper is as follows:
Sec. \ref{silent} provides a derivation of the silent universe;
Sec. \ref{code} describes the code \textsl{simsilun};
Sec. \ref{results} presents the results of the Simsilun simulation and shows that a natural consequence of the relativistic evolution is the emergence of spatial curvature  $\Omega_K$, and slightly lower values $\Omega_M$ and $\Omega_\Lambda$ compared to the $\Lambda$CDM model; Sec. \ref{conclusions} presents the conclusions.
}

\section{`Silent' approach to relativistic cosmology}\label{silent}

\subsection{Relativistic evolution of a cosmic fluid}

Below the equations for the silent universe are derived. 
The material is compiled from the work by \citet{1971grc..conf..104E,2009GReGr..41..581E},
\citet{1993GReGr..25.1225E}, \citet{1997CQGra..14.1151V}, 
\citet{2006igrc.book.....P}, and \citet{2008PhR...465...61T}.

{  
The silent universe and its evolutionary equations are derived using the  1+3 split 
(time+space) and comoving gauge. Within this approach one assumes existence a unique vector filed $u^a$ that can be associated with the flow of matter (this could be an average velocity field of a given domain). Then, worldlines of particular domains (and time) are defined as lines that are tangent to $u^a$, and space is defined as a surface orthogonal to $u^a$. }
A cosmological fluid with its velocity $u^a$ can then be  described with the following energy momentum tensor

\begin{equation}
T_{ab}= \rho u_a u_b+ p h_{ab}+ \pi_{ab} + 2q_{(a} u_{b)},
\end{equation}
where $\rho$ is energy density, $p$ is pressure, $\pi_{ab}$ is the anisotropic stress tensor ($\pi_{ab} u ^b = 0$),
$q^a$ is the heat-flow vector ($q_a u^a = 0$), and $h_{ab}$ is the spatial part of the metric in 3+1 split $h_{ab}= g_{ab} - u_a u_b$. Introducing the following notation
$ \dot{X}_{ab...} = u^c \nabla_c X_{ab..}$ and $D_c X_{ab..} = h_c{}^m h_a{}^n h_b{}^p \nabla_m X_{np}$, the gradient of the velocity field can be written as
\begin{equation}
 u_{a;b} = \omega_{a b} + \sigma_{a b}
+ \frac{1}{3} h_{a b} \Theta - A_a u_b, 
\end{equation}
where
$\omega_{a b} = D_{[b} u_{a]}$ is rotation, $\sigma_{a b} = D_{\langle b} u_{a \rangle}$ is shear (the angle bracket is the projected symmetric trace-free part: $S_{\langle a b \rangle} = S_{(ab)} - S h_{ab}/3)$, $\Theta = D^a u_a$ is expansion, and $ A^a = \dot{u}_a$ is acceleration.
The evolution of the fluid follows from the conservation equations $T^{ab}{}_{;b} =0$ 
 \begin{eqnarray}
&& \dot{\rho}  + \Theta(\rho+p) + \sigma^{ab}\pi_{ab} + q^a{}_{;a} + q^a A_a = 0, \label{ffe1} \\   
&&  (\rho+p)A_a +  {\rm D}_ap +  {\rm D}^b\pi_{ab} + \pi_{ab}A^b  + (\omega_a{}^b + \sigma_a{}^b + \frac{4}{3} \Theta h_a{}^b) q_b = 0,
 \end{eqnarray}
and the evolution of the velocity field follows from the Ricci identities ($u_{a;d;c}-u_{a;c;d} = R_{abcd} u^b$) 

\begin{eqnarray}
&& \dot{\Theta}  = - \frac{1}{3}\,\Theta^2-  \frac{1}{2}\,(\rho+3p)-
2(\sigma^2-\omega^2)+ {\rm D}^aA_a  + A_aA^a+ \Lambda,  \\
&&  \dot{\sigma}_{\langle ab\rangle}  = - \frac{2}{3}\,\Theta\sigma_{ab}-
\sigma_{c\langle a}\sigma^c{}_{b\rangle}- \omega_{\langle
a}\omega_{b\rangle}  + {\rm D}_{\langle a}A_{b\rangle}+ A_{\langle
a}A_{b\rangle} - E_{ab}+  \frac{1}{2}\,\pi_{ab},\\
&& \dot{\omega}_{\langle a\rangle}  = - \frac{2}{3}\,\Theta\omega_a-
 \frac{1}{2}\,{\rm curl} \, A_a+ \sigma_{ab}\omega^b,
 \end{eqnarray}
where ${\rm curl} \, S_{ab} = \epsilon_{ad(a}D^c S_{b)}^d$,  $\omega_a = \epsilon_{abc} \omega^{bc}/2$, and $E_{ab}$ is the electric part of the Weyl tensor $C_{acbd}$, which together with the magnetic part is defined as
$ E_{ab} = C_{acbd} u^c u^d$, and $H_{ab} = \frac{1}{2} \epsilon_a{}^{cd} C_{cdbe} u^e$.
The evolution of the Weyl curvature follows from the 
Bianchi identities ($R_{ab[cd;e]} = 0$) 
\begin{eqnarray}
&&  \dot{E}_{\langle ab\rangle}  = -\Theta E_{ab}-
 \frac{1}{2}\,(\rho+p)\sigma_{ab}+  {\rm curl} \, H_{ab}-
 \frac{1}{2}\,\dot{\pi}_{ab}  -  \frac{1}{6}\,\Theta\pi_{ab} \nonumber \\
&& + 3\sigma_{\langle
a}{}^c\left(E_{b\rangle c}- \frac{1}{6}\,\pi_{b\rangle c}\right)  + \varepsilon_{cd\langle
a}\left[2A^cH_{b\rangle}{}^d-\omega^c\left(E_{b\rangle}{}^d+
 \frac{1}{2}\,\pi_{b\rangle}{}^d\right)\right],   \\
&&  \dot{H}_{\langle ab\rangle} = -\Theta H_{ab}- {\rm curl} \, E_{ab}+
 \frac{1}{2}\,{\rm curl} \, \pi_{ab}  +3\sigma_{\langle a}{}^cH_{b\rangle c}
-\varepsilon_{cd\langle a}\left(2A^cE_{b\rangle}{}^d+\omega^cH_{b\rangle}{}^d\right). \nonumber \\
&& \label{ffe10} 
\end{eqnarray}

The above equations provide the description of the evolution of a cosmic fluid. Thus, the state of the cosmic fluid is defined by properties of the fluid  (density $\rho$, pressure $p$ and $\pi_{ab}$, and energy transfer $q_a$), its velocity filed (expansion rate $\Theta$, shear $\sigma_{ab}$, rotation $\omega_{ab}$, and acceleration $A_a$), and spacetime curvature (electric $E_{ab}$ and magnetic $H_{ab}$ parts of the Weyl tensor).
In order to evolve the fluid (i.e. solve eqs. (\ref{ffe1})--(\ref{ffe10})) one also needs to specify the initial conditions and satisfy the spatial constraints, i.e. spatial parts of the Ricci identities 

\begin{eqnarray}
&& D^b \sigma_{ab} = \frac{2}{3} D_a \Theta + {\rm curl} \, \omega_a + 2 \epsilon_{abc}A^b \omega^c - q_a, \label{brc1} \\
&& D^a \omega_a = A_a \omega^a, \label{brc2}  \\
&& H_{ab} = {\rm curl} \, \sigma_{ab} + D_{\langle a} \omega_{b \rangle} + 2 A_{\langle a} \omega_{b \rangle}, \label{brc3} 
\end{eqnarray}
and Bianchi identities 
\begin{eqnarray}
&& D^b E_{ab} = \frac{1}{3} D_a \rho - \frac{1}{2} D^b \pi_{ab} - \frac{1}{3} \Theta q_a + \frac{1}{2} \sigma_{ab} q^b - 3 H_{ab}\omega^{b} + \epsilon_{abc} (\sigma^b{}_{d}H^{cd} - \frac{3}{2} \omega^b q^c), \nonumber \\ && {} \label{brc4} \\
&& D^b H_{ab} = (\rho+p) \omega_a - \frac{1}{2} {\rm curl}\, q_a -  \frac{1}{2} \pi_{ab} \omega^b + 3 E_{ab}\omega^{b} - \epsilon_{abc} \sigma^b{}_{d} (E^{cd} + \frac{1}{2} \pi^{cd}). \nonumber \\ && {} \label{brc5}
\end{eqnarray}

\subsection{Evolution of the irrotational silent universe}

Irrotational silent universe is such a solution of the above equations, where each worldline 
evolves independently of other worldlines, and apart from the initial constraints there is no communication between the worldlines, i.e. no pressure gradients, no energy flux, no gravitational radiation:

\[\omega_{ab} = 0, \quad A_{a} = 0, \quad q_a = 0, \quad  D_a p=0, \quad \pi_{ab} = 0, \quad H_{ab} = 0.\]
\noindent These constraints have strong implications. For example from Bianchi and Ricci identities with $H_{ab} =0$ follows that 
\[ {\rm curl} \, \sigma_{ab} = 0  \quad {\rm and} \quad \epsilon_{abc} \, \sigma^b{}_d E^{cd} = 0,\]
which means that inhomogeneous models
with diagonalisable 
the shear and electric part of the Weyl tensor
have  only 1 independent component of the shear and only 1 independent component of the electric part of the Weyl tensor (to be precise: 2 eigenvalues are identical, and the third is fixed by the  condition of vanishing trace). This implies that the Petrov type I models are ruled out (for more details see a rigorous derivation by \citet{1997CQGra..14.1151V}). Thus the most general silent inhomogeneous models are Petrov type D, and so the shear tensor and the electric Weyl tensor can be written as

\begin{equation}
\sigma_{ab} = \Sigma \, {\rm e}_{ab}, \quad E_{ab} = {\cal W} \, {\rm e}_{ab}, \label{brcs}
\end{equation}
where ${\rm e}_{ab} = h_{ab} - 3 z_a z_b$ where $z^a$ is a space-like unit vector aligned with the Weyl principal tetrad. As a result the fluid equations (\ref{ffe1})--(\ref{ffe10})
reduce only to 4 scalar equations \citep{1995ApJ...445..958B,1997CQGra..14.1151V} 

\begin{eqnarray}
&& \dot \rho = -\rho\,\Theta, \label{rhot}\\
&& \dot \Theta = -\frac{1}{3}\Theta^2-\frac{1}{2}\,\kappa \rho-6\,\Sigma^2 + \Lambda,\label{thtt}\\
&& \dot \Sigma = -\frac{2}{3}\Theta\,\Sigma+\Sigma^2-{\cal W},\label{shrt}\\
&& \dot{ {\cal W}} = -\Theta\, {\cal W} -\frac{1}{2}\,\kappa \rho\,\Sigma-3\Sigma\,{\cal W}.\label{weyt}
\end{eqnarray}

It is interesting to note that initially it was thought that the evolution of the silent universe would be described by 6 scalars: 1 density, 1 expansion rate, 2 for the shear tensor, and 2 for the electric part of the Weyl tensor \citep{1995ApJ...445..958B}, but then it was proved by \citet{1997CQGra..14.1151V} and also by \citet{1997PhRvD..55.5219M}
that the constraints (\ref{brc3}) and (\ref{brc5}) reduce the number of scalars to 4. 
Recently, \citet{SzekInterUJ} showed that for 4 scalars  the evolution equations are completely integrable, but in the case of 6 scalars the system is not integrable, i.e. the Darboux polynomials method yields only 1 first integral.

\subsection{Initial conditions, spatial constraints, and the approximation of the \textsl{Simplified Silent Universe}}

When specifying the initial conditions one needs to make sure that the spatial constraints  (\ref{brc1})--(\ref{brc5}) are satisfied. If these constraints are initially satisfied, they will be preserved in the course of the evolution \citep{1997CQGra..14.1151V,1997PhRvD..55.5219M}.
These conditions involve covariant derivatives so in order to satisfy them one needs to
know the Christoffel symbols, which usually implies knowledge of the metric. It is possible to replace the terms that involve covariant derivatives with other quantities, consequently replacing them with algebraic relations \citep{2012CQGra..29f5018S}, but eventually one either needs to know the metric to evaluate these terms or use some other algorithm to constrain these quantities. Below, such an algorithm is presented, and it is based on several approximations.
We first start with writing down the perturbations for the density filed

\begin{equation} 
\rho = \bar{\rho} + \Delta \rho = \bar{\rho} \, ( 1 + \delta), 
\end{equation}
where  $\bar{\rho}$ is the background density, and $\delta$ is the density contrast.
In the early universe matter density is much larger than the contribution from the cosmological constant, and so the universe (or the $\Lambda$CDM model, which is used in this paper) evolves as the Einstein-de Sitter model. Assuming that the perturbations are small and dominated by the growing mode  \citep{1980lssu.book.....P}  the expansion rate is
\begin{equation}
\Theta = \bar{\Theta} + \Delta \Theta =  \bar{\Theta} \, ( 1 - \frac{1}{3}  \, \delta).
\end{equation}
Inserting (\ref{brcs}) to  (\ref{brc1}) and (\ref{brc4}), and assuming that the early universe is well approximated by the Einstein-de Sitter background (i.e. $3\bar{\rho} = \bar{\Theta}^2 $) we get that
\begin{eqnarray}
&& {\rm e}_{ab} \, D^b \Sigma + \Sigma \, D^b {\rm e}_{ab} = \frac{2}{3} D_a \Theta  = - \frac{2}{9} \bar{\Theta}^2 D_a \delta, \\
&& {\rm e}_{ab} \, D^b {\cal W} + {\cal W} D^b {\rm e}_{ab} = \frac{1}{3} D_a \rho =  
\frac{1}{3} \bar{\rho} D_a \delta_i  = \frac{1}{9} \bar{\Theta}^2   D_a \delta,
\end{eqnarray}
where above it was also assumed that the shear $\Sigma$ is a perturbation, so $\Sigma \Theta \approx \Sigma \bar{\Theta}$.
Comparing the right hand sides of the above equations and neglecting higher order terms (such as $\Sigma \, \delta$), the above equations reduce to

\begin{equation} 
{\cal W} =  - \frac{1}{2} \bar{\Theta} \, \Sigma =  - \frac{3}{2}  \frac{\Sigma}{\bar{\Theta}} \, \bar{\rho}.
\end{equation}
Up to this stage, the explicit knowledge of the metric was not essential. However, in order to find the relation for ${\cal W}$ or $\Sigma$, which is needed to provide all 4 initial conditions, one needs to know the metric. Without the explicit knowledge of the metric, one can only rely on further approximations. Here we are going to apply an approximation that is based on the exact formula for the quasi-local perturbations and averages \citep{2012CQGra..29f5018S}, which is

\begin{equation} 
\Sigma = -\frac{1}{3} \, \Delta \Theta. 
\end{equation}
Although this relation is exact for the quasi-local quantities, it is not the exact relation for local quantities. 
For local quantities this relation follows if one assumes that locally the spacetime is flat ($\nabla_a \approx \partial_a$), then one can integrate eq. (\ref{brc1}) and arrive at the above formula.
This allows us to close all the equations and write down the formulae for the initial conditions

\begin{eqnarray}
&& \rho_i = \bar{\rho} + \Delta \rho = \bar{\rho} \, ( 1 + \delta_i), \label{rhoi} \\
&& \Theta_i = \bar{\Theta} + \Delta \Theta =  \bar{\Theta} \, ( 1 - \frac{1}{3}  \delta_i), \label{thti} \\
&& \Sigma_i = - \frac{1}{3} \, \Delta \Theta = \frac{1}{9}  \bar{\Theta} \, \delta_i, \label{sigmai} \\
&& {\cal W}_i = - \frac{1}{6} \bar{\rho} \, \delta_i, \label{weyi}
\end{eqnarray}
where the subscript $i$ denotes initial values, and $\delta_i$ is the initial density contrast.
The first equations is just a definition of a perturbation $\delta$. The second one is exact as long as the perturbations are small and the background is well approximated by the Einstein-de Sitter
model (this is satisfied in the early universe). The relation between the shear $\Sigma$ and the Weyl curvature ${\cal W}$ is also based on the linear perturbations around the Einstein-de Sitter model.
However, eq. (\ref{sigmai}) is more than just a linear approximation. This is an exact relation for quasi-local quantities \citep{2012CQGra..29f5018S} but for local quantities it is an approximation. This approximation reduces the richness of all possible initial conditions that enter via (\ref{brc1}) to a single quantity $\delta$. 
Therefore, the above set of initial conditions is referred to as the \textsl{Simplified Silent Universe}, and the code that implements these relations is named \textsl{simsilun}.

\subsection{Comparison with the Szekeres model}\label{szcomp}

The Szekeres model is the most general, explicit solution that belongs to the class of the silent universes. Therefore, it can be used to test the above approximations, which form the basis of the \textsl{Simplified Silent Universe}.
The metric of the quasispherical Szekeres model is usually represented in the following form 
\citep{1996CQGra..13.2537H}

\begin{equation}
ds^2 =  c^2 dt^2 - \frac{\left(R' - R \frac{{E}'}{{ E}}\right)^2}
{1 - K} dr^2 - \frac{R^2}{{  E}^2} (dp^2 + dq^2), \label{ds2}
 \end{equation}
where ${}' \equiv \partial/\partial r$, $R = R(t,r)$,  and $K = K(r)$ is an arbitrary function of $r$. The function ${  E}$ is given by
 \begin{equation}
{  E}(r,p,q) = \frac{1}{2S}(p^2 + q^2) - \frac{P}{S} p - \frac{Q}{S} q + \frac{P^2}{2S} + \frac{Q^2}{2S} + \frac{S}{2}  ,
 \end{equation}
where the functions $S = S(r)$, $P = P(r)$, $Q = Q(r)$.
The evolution of the system follows from the Einstein equations and reduces to a single equation 
\begin{equation}\label{evo}
\dot{R}^2 = -K(r) + \frac {2 M(r)} {R} + \frac 1 3 \Lambda R^2,
\end{equation}
where $\dot{} \equiv \partial/\partial t$,
$\Lambda$ is the cosmological constant, and $M(r)$ is an arbitrary
function. The above equation can be integrated and used to define one more arbitrary function, i.e. the bang time function $t_B(r)$

\begin{equation}
t - t_B(r) = \int\limits_0^{R}\frac{{\rm d} \widetilde{R}}{\sqrt{- K +
2M / \widetilde{R} + \frac 1 3 \Lambda  \widetilde{R}^2}}.
\label{tbf}
\end{equation}
The fluid scalars: density $\rho$, expansion $\Theta$, shear $\Sigma$, and Weyl curvature ${\cal W}$ can be expressed in terms of the function $R(t,r)$ and the arbitrary functions $E(r)$, and $M(r)$ \citep{2002PhRvD..66h4011H} 

\begin{eqnarray}
&& \rho =  \frac {2 \left(M' - 3 M  E' /  E\right)} {R^2 \left(R' - R E' /  E\right)},\label{szrho} \\
&& \Theta = \frac{ \dot{R}' + 2 \dot{R}R'/R - 3 \dot{R}  E' /  E}{R' - R E' /  E}, \label{sztht} \\
&& \Sigma = -\frac{1}{3}  \frac{ \dot{R}' - \dot{R}R'/R}{R' - R E' /  E},  \label{szshr} \\
&& {\cal W} =  \frac{M}{3R^3} \frac{3R' - R M'/M} { R' - R E' /  E}. \label{szwey}
\end{eqnarray}

\begin{figure}
\begin{center}
\includegraphics[scale=0.7]{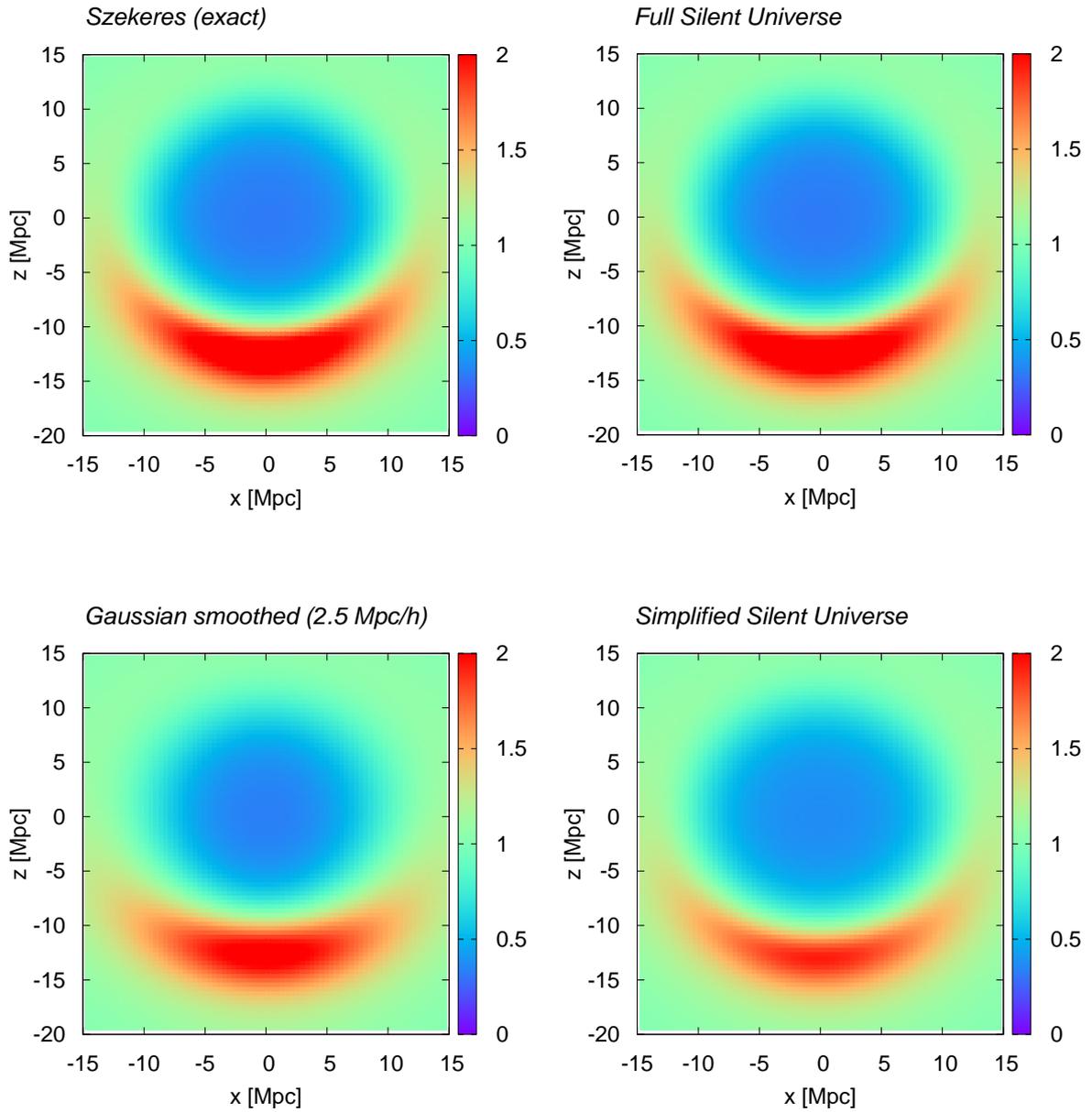}
\end{center}
\caption{Present-day density distribution normalised by the density of the $\Lambda$CDM model ($\rho/\rho_{\Lambda CDM}$). {\em Upper Left \/} panel shows the density distribution obtained within the Szekeres model, and {\em Upper Right} within the Full Silent Universe (these two approaches produce identical results); {\em Lower Left} panel shows the density distribution of the Szekeres model smoothed with the Gaussian kernel of $2.5 h^{-1}$ Mpc, and {\em Lower Right} panel shows the density distribution obtained within the Simplified Silent Universe.}
\label{fig1}
\end{figure}

Thus, to calculate the evolution of the system and evaluate the fluid scalars (density $\rho$, expansion $\Theta$, shear $\Sigma$, and Weyl curvature ${\cal W}$) one needs to know the form of the arbitrary functions $S$, $P$, and $Q$  (to get $E'/E$), and $M$ and $K$ to get the evolution of $R$. For the purpose of this Section, the function $M$ is set to be
\begin{equation}
M(r) = \frac{1}{6} 8 \pi G \rho_{CMB} \left[ 1 + \frac{1}{2} m_0 \left( 1 - \tanh \frac{r-r_0}{2 \Delta r} \right)  \right] r^3, 
\end{equation}
$m_0 = -0.002$, $ \Delta r = r_0/4.0$, and $r_0 = 25.0$,
$\rho_{CMB} = (1+z_{CMB})^3 \Omega_M 3 H_0^2/(8 \pi G)$ and $z_{CMB} = 1090.0$ and 
 $\Omega_M = 0.308$, and $H_0 = 67.81$ km s$^{-1}$ Mpc$^{-1}$, and
the cosmological constant is set to be $\Lambda = \Omega_\Lambda 3 H_0^2$, where  $\Omega_\Lambda = 0.692$ \citep{2016A&A...594A..13P}. The function $K$ is fixed by the condition of a uniform age of the universe $t_B = 0$, so that it corresponds to existence of pure  growing modes \citep{2009suem.book.....B,2013CQGra..30w5001S}. 
Finally, the functions that define the dipole $E'/E$ are 
\begin{eqnarray}
&& S(r)  = r^{2/5},  \nonumber \\
&& P(r) = 0,  \nonumber \\
&& Q(r) = 0.
\end{eqnarray}

The model is specified at the last scattering instant and its evolution is calculated up to the present instant by integrating eq. (\ref{evo}). 
The present-day density distribution is presented in the upper left panel of Fig. \ref{fig1}.
Then, at the initial instant, the initial values for $\rho$, $\Theta$, $\Sigma$, and ${\cal W}$ 
were obtained from eqs. (\ref{szrho})--(\ref{szwey}). These initial conditions were then used to 
calculate the evolution of the system (\ref{rhot})--(\ref{weyt}), and the resulted present-day density distribution is presented in the upper right  panel of Fig. \ref{fig1}.
The numerical solution of eqs. (\ref{rhot})--(\ref{weyt})  with the initial conditions given by eqs. 
(\ref{rhoi})--(\ref{weyi}), which corresponds to Simplified Silent Universe is presented in the lower right panel of Fig. \ref{fig1}. For comparison, the 
Gaussian smoothed density distribution of the Szekeres model (upper left) is presented in the lower left panel of  Fig. \ref{fig1}.
As seen, there is a reasonably good agreement between the full solution and the simplified solutions, which justifies the application of the Simplified Silent Universe.

\section{The code: \textsl{simsilun}} \label{code}

\begin{figure}
\begin{center}
\includegraphics[scale=1.0]{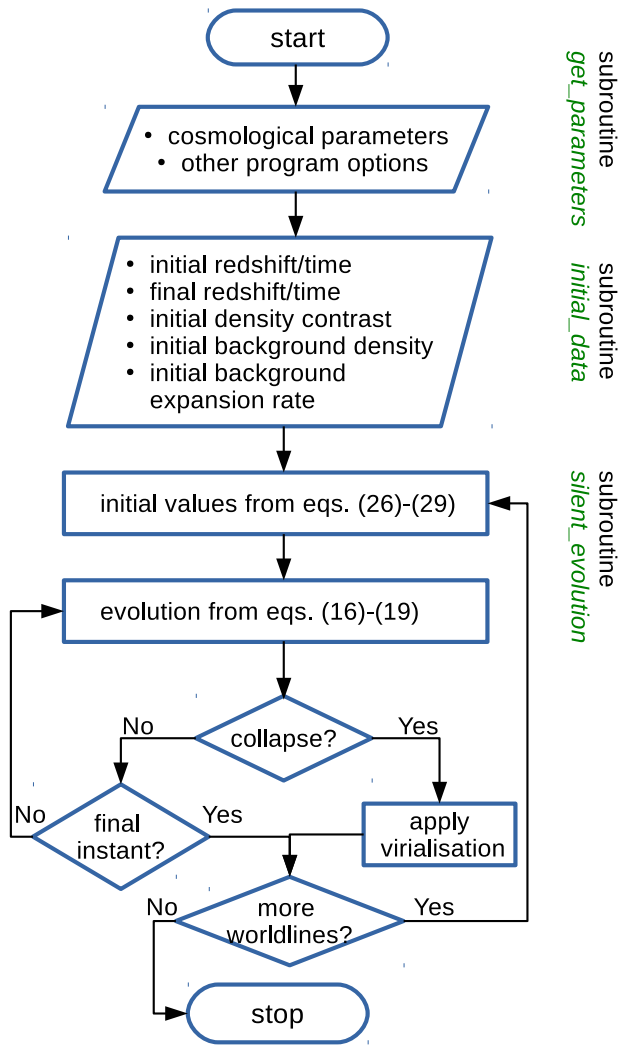}
\end{center}
\caption{Algorithm's flowchart. See Sec. \ref{code} for details.}
\label{fig2}
\end{figure}

The code is available via the \textsl{Bitbucket repository}\footnote{\url{https://bitbucket.org/bolejko/simsilun}}. The code is written in Fortran. The algorithm on which the code is based is presented in Fig. \ref{fig2}.
The code consists of 3 subroutines: \textsl{get\_parameters} (the subroutine that contains the cosmological parameters and other model parameters), \textsl{initial\_data} (the subroutine that reads the initial data such as: initial and final redshift/instant, initial background's density and expansion rate, and initial density contrast), and \textsl{silent\_evolution} (the subroutine that calculates the evolution). The evolution is calculated by numerically solving eqs. (\ref{rhot})--(\ref{weyt}) with the 4th order Runge--Kutta method, and the initial conditions given by  eqs. (\ref{rhoi})--(\ref{weyi}).

Within the silent universe, vorticity and pressure gradients are absent. Consequently, 
once the density contrast is too large and matter starts to collapse, there is nothing that could prevent the singularity. In order to prevent the singularity, some virialisation mechanism needs to be implemented. Since the model lacks the rotation and gradients, the virialisation needs to be implemented externally. 
In the code \textsl{simsilun} three types of virialisation mechanisms are implemented:

\begin{itemize}
\item Stop at the turnaround

The code stops at the turnaround, i.e. when $\Theta$ becomes negative. The 
fluid scalars are recoded as they are at this instant, and the expansion rate is set to zero, $\Theta =0$. The code then  assumes that these quantities are `frozen' and do not change across the cosmic evolution any more. 

\item Stop near the singularity

The code stops at the singularity, and restores the 
fluid scalars as they were 2 numerical steps before the singularity\footnote{Since the code reaches the singularity, users are advised to check how their compilers handle it. The code was tested with \textsl{ifort} and \textsl{gfortran} and these compilers are able to handle NaN and restore the fluid scalars as they were before the singularity.}. These values are then recoded, the expansion rate is set to zero, $\Theta =0$, and the code  assumes that these quantities do not change any more. 

\item Virialised halo

Here it is assumed that the end point of a collapse is a virialised halo with the NFW profile. The volume of the halo is  $V_{halo} =  M/(\bar{\rho}(t) \Delta) $,  where $\bar{\rho}(t)$ is matter density of the background model (here it is the $\Lambda$CDM model), $\Delta = 180$, and $M$ is the mass \citep{1996ApJ...462..563N,2001MNRAS.321..372J}. Since the mass is conserved throughout the evolution its value is the same as at the initial instant, $M_i= \rho_i \, V_i$,  where $\rho_i$ is the initial density and $V_i$ initial volume.
Matter density evolves with time, $\bar{\rho}(t)$, so the volume occupied by the virialised halo also evolves with time. 
The code stops and the turnaround, and replaces the collapsing region with a virialised halo.
To account for the transition zone between the virialised halo and the surrounding universe it is assumed that the volume of this element is $ V = 3 \, V_{halo} = 3 M/(\bar{\rho}(t) \Delta) $, and so the density of the cell is 
$\rho = \bar{\rho}(t) \Delta /3$. The other fluid scalars are assumed to be zero, ie. $\Theta =0 = \Sigma = {\cal W}$.
\end{itemize}

The code can be run as it is stands, without any modifications. However, the user is encouraged to modify the code to meet any specific needs. For example, in the code the initial data is specified as a vector of 2000 different worldliness. This part can easily be modified by the user to include any other data (as for example in Sec. \ref{results}). Similarly, the initial conditions for $\rho$, $\Theta$, $\Sigma$, and ${\cal W}$ are based on the conditions for the \textsl{Simplified Silent Universe} (\ref{rhoi})--(\ref{weyi}). This part can also be modified by the user to include the full exact solution of (\ref{brc1}) and (\ref{brc4}) (as for example in Sec. \ref{szcomp}).

\begin{figure}
\begin{center}
\includegraphics[scale=1.0]{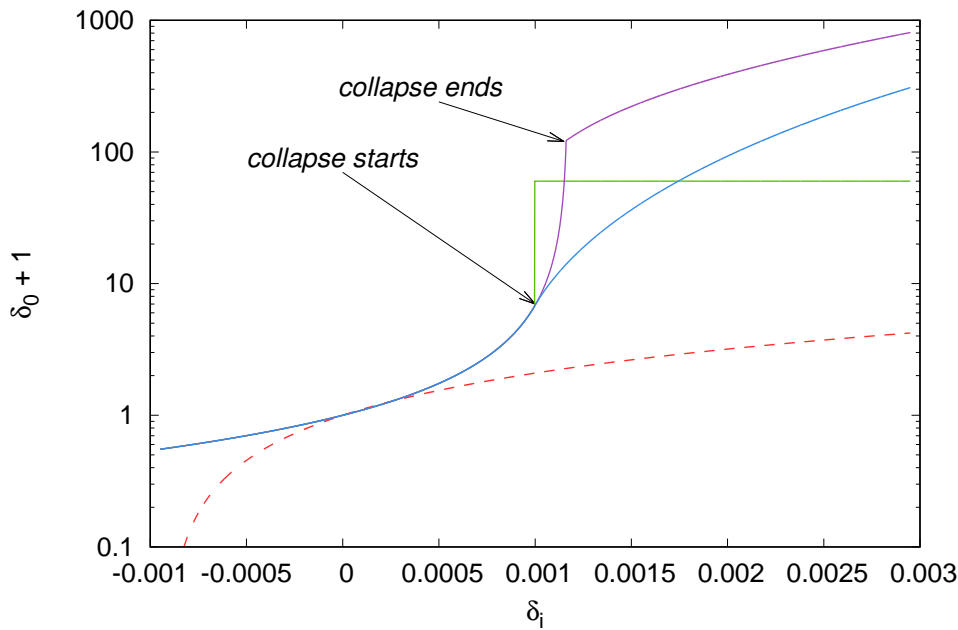}
\end{center}
\caption{Present day density ($\delta_0+1$) obtained from the code \textsl{simsilun} as a function of the initial density contrast ($\delta_i$).  After compiling and running the code (see the file \textsl{readme} for instructions) the code should produce a similar figure. It will show one solid line (depending on which virialisation scenario is implemented) and for comparison it will also show the density contrast evaluated based on linear perturbations around the Einstein-de Sitter background (dashed line).}
\label{fig3}
\end{figure}

After compiling and running the code (see the file \textsl{readme} for instructions) the code produces results presented in Fig. \ref{fig3}. Figure \ref{fig3} shows the present-day density (to be precise it is $\delta_0+1$ so that it shows well on a log-scale plot) obtained from the code. It is assumed that the initial instant is at the last scattering $z=1090$, and that the present instant is $z=0$. Three different lines show three different virialisation scenarios, and for comparison the dashed line shows the density contrast evaluated based on linear perturbations around the Einstein-de Sitter
 background\footnote{  {  The evolution of the linear density contrast around the Einstein-de Sitter
model is for comparison only. The evolution of the linear density contrast is approximated with $\delta_{EdS} = \delta_i \, (\rho/\rho_i)^{1/3}$, which is exact for the Einstein-de Sitter model. 
For other FLRW backgrounds the linear density contrast follows different evolution, which can be up to a factor of 2 different from the evolution of the linear density contrast within the Einstein-de Sitter model.  } 
}.

For initial density contrasts $\delta_i < 0.001$ the present day regions are still expanding and so no virialisation mechanism needs to be implemented. For large values of $\delta_i$, at the present day, when $\delta_0 \approx 5.75$ the region starts to collapse and $\Theta < 0$. For even larger $\delta_i$ the turnaround starts earlier, before the present-day instant. 
At this instant either virialisation scenario 1 or scenario 3 (depending on the option set in the subroutine \textsl{get\_parameters}) activates. Otherwise the region collapses up to the singularity when the scenario 2 is activated; 
at the present-day instant this happens for $\delta_i \approx 0.0012$ for large values of $\delta_i$, this stage is reached earlier, before the present day.

\subsection{Limitations and approximations of the Simplified Silent Universe}

There are two major limitations and approximations that are at the core of the Simplified Silent Universe. Firstly, it is assumed that locally the evolution of the universe can be approximated with the silent universe. Secondly, the virialisation is implemented externally.

The first limitation, boils down to applicability  of eqs. (\ref{rhot})--(\ref{weyt}).
These equations break down once the shear tensor can no longer be diagonalised.
This happens at small scales, where the accretion and cosmic flows have a complicated geometry,
and evolution becomes highly non-linear ---  the assumption of the silent universe are 
preserved by linear and second order perturbations \citep{2002PhRvD..66l4015E}.
The assumption of the silent evolution leads to a system, where each worldline 
evolves independently from other worldlines (eqs. (\ref{rhot})--(\ref{weyt}) do not contain gradients).
However, this does not mean that worldlines are completely independent from each other.
The dependence enters via the spatial constraints  (\ref{brc1})--(\ref{brc5}).
Within the silent universe, if these constraints are initially satisfied, they
will be preserved in the course of the evolution \citep{1997CQGra..14.1151V,1997PhRvD..55.5219M}.
These constraints are actually what makes various exact solutions of the Einstein equations 
so different from each other. There are number of exact solutions, whose evolution is govern
by eqs. (\ref{rhot})--(\ref{weyt}), for example Szekeres models, the LTB models, some Bianchi models,
and some LRS models. So although they are govern by the same evolutionary equations,
they obey different spatial constraints. This makes some cosmologists wonder if there are other yet-to-be-discovered exact solutions that are silent. 
Such a solution would be more general than the known solutions, but 
its evolution would also be govern by eqs. (\ref{rhot})--(\ref{weyt}). 
The idea behind the Simplified Silent Universe 
is based on the assumption that such a more general solution exists, i.e. there exists a solution that  obeys eqs. (\ref{rhot})--(I\ref{weyt}) and can be initialised with a reasonable and cosmologically justified set of initial conditions. In Sec. \ref{results} we will use the Millennium Simulation to set up the initial conditions.

The second limitation is that one needs to externally implement some sort of virialisation mechanism.
Once the evolution becomes highly nonlinear the silent universe collapses into a singularity, whereas in the real universe such a system would undergo virialisation.
In the code \textsl{simsilun} three types of virialisation mechanisms are implemented.
{  In the literature of inhomogeneous cosmological models, other types of virialisation include pressure gradients \citep{2008MNRAS.391L..59B}, and variants of virialisation scenarios 2 and 3 \citep{2012JCAP...05..003B,2013JCAP...10..043R,2017arXiv170606179R}.
Also, what is beyond the scope of the present work is studying
the transition from the pre-virialisation to post-virialisation stages,
analysed by \citet{2017arXiv170606179R}. However, as will be shown in Sec. \ref{results}, the global results do not significantly depend on the assumed type of the virialisation mechanism. 
}

The above simplifications and approximations of the Simplified Silent Universe limit its applicability.
However, it is expected that on scales beyond 2-5 Mpc, and in the mildly nonlinear regime, the 
Simplified Silent Universe should work well.
Comparisons with other approaches to numerical cosmology, such as the ones that are based on the BSSN formalism will show how well the above approximations work.
The advantage of these approximations and simplifications is that one can easily model complicated cosmological systems and trace their evolution without extensive CPU calculations.
In addition, exploring the evolution of the universe with the Simplified Silent Universe
allows to study features that are absent within the $\Lambda$CDM model
such as position-dependent expansion rate and the emergence of spatial curvature.
Understanding these two processes is a key element in understanding the phenomenon of backreaction.

\section{Relativistic modelling of the evolution of the large scale structure of the universe}\label{results}

\subsection{Using the Millennium Simulation to set up the initial conditions for the Simsilun simulation}\label{mfield}

\begin{figure*}
\begin{center}
\includegraphics[scale=0.65]{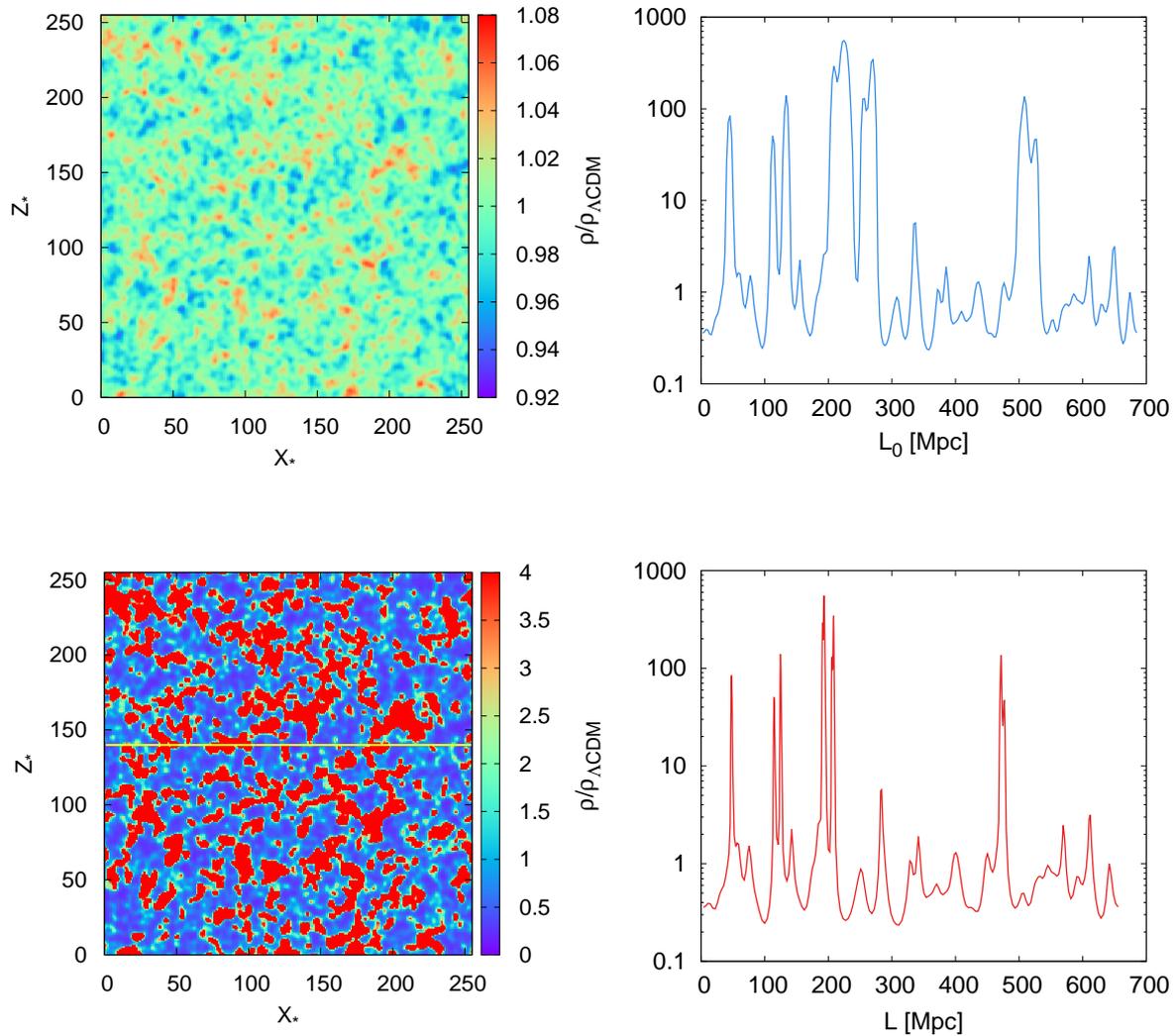}
\end{center}
\caption{ {\em Upper left:} initial density distribution across the cells with $Y_* = 120$ (i.e. out of $256^3$ cells, only cells with $Y_* = 120$ are presented). {\em Lower left:} present-day density distribution across the same cells. {\em Upper right:} present-day density profile along the cells with $Y_* = 120$ and $Z_* = 140$, the length $L_0$ represents a uniform conversion (as in the Millennium simulation) where each cell has a size of 2.68 Mpc. {\em Lower right:} present-day density profile along the cells with $Y_* = 120$ and $Z_* = 140$, with the physical length ($L$) of the structures.}
\label{fig4}
\end{figure*}

The code \textsl{simsilun} evaluates the evolution of the universe based only on the initial density contrast. 
In this section we use the Millennium Simulation \citep{2005Natur.435..629S,2009MNRAS.398.1150B,2013MNRAS.428.1351G}. We use the smoothed matter density field, which is stored in the MField database and accessible via an online service provided by the German Astrophysical Virtual Observatory\footnote{The GAVO portal is at \url{
http://gavo.mpa-garching.mpg.de/MyMillennium} and requires registration in order to access the data. The SQL query used to access the MField is: \texttt{select * from MField..MField}.} \citep{2006astro.ph..8019L}.
The cosmological parameters of the Millennium simulations are based on WMAP1, i.e.  $\Omega_M = 0.25$, $\Omega_\Lambda = 0.75$, $H_0 = 73.0$ km s$^{-1}$ Mpc$^{-1}$,
{  and $\sigma_8 =  0.9$}.
 These cosmological parameters were updated in the subroutine  \textsl{get\_parameters}. Also, the subroutine \textsl{initial\_data} was modified to read the MField data. Finally, the initial instant was set to $z=79.997894$, which corresponds to the first snapshot of the Millennium simulation.
The MField consists of $256^3$ cells (grids). Each cell contains information about the smoothed density field. Here we use the density field smoothed with a radius $2.5 \, h^{-1}$ Mpc {  . The smoothing reduces the variance of the density filed, which if measure by the parameter $\sigma_8$ is reduced by approximately 10\% compared to underling value of $\sigma_8$}.
The MField data was used as the initial condition for the code \textsl{simsilun}, and the resulting simulation is referred to as the Simsilun simulation.

The Simsilun simulation consists of 16,777,216 worldlines, which are evolved from $z=79.997894$ to $z = 0$.
{  The variance of the density filed, at the present day ($z=0$), if averaged over the spheres of radius $8 \, h^{-1}$ Mpc is $0.78$. This value is lower than the 
parameter $\sigma_8$ of the Millennium simulation, but comparable to the parameter $\sigma_8$ evaluated over the MFiled smoothed with $2.5 \, h^{-1}$ Mpc.}
The initial density distribution is presented in the upper left panel in Fig. \ref{fig4}. This panel shows a slice through the simulation with a  plane of $Y_* = 120$, i.e. out of $256^3$ cells, only cells with $Y_* = 120$ are presented.
The lower left panel in Fig. \ref{fig4} shows the density distribution at the present instant. Again, this is the density distribution across the cells with $Y_* = 120$. It should be noted, that the distribution of cells is not the same as the actual distribution of matter. This is due to the fact that cells do not evolve uniformly. In the Millennium simulations each cell evolves in the same way, and at the present day each cell has the same size of 2.68 Mpc ($500.0 \times 0.73^{-1} \times 256^{-1}$ Mpc $= 2.68$ Mpc). Using this conversion (hence the name  $L_0$), the present-day density profile along a slice of $Y_* = 120$  and $Z_* = 140$ (this line is also depicted in the lower right panel) is presented in the upper right panel in Fig. \ref{fig4}. However, the real profile along this line (the axis is named $L$ to distinguish between the Millennium's $L_0$ and cells' $X_*$)
is  presented in the lower right panel in Fig. \ref{fig4}. As seen, overdensities are in fact smaller and underdense regions are larger, compared to the grid (cell) representation. Thus, the real distribution of structures is not the same as the one presented in 
 lower left panel in Fig. \ref{fig4}, which distorts the real picture (making overdense regions look larger and underdense regions look smaller).

\subsection{Evolution of matter, dark energy and spatial curvature}

\begin{figure}
\begin{center}
\includegraphics[scale=1.0]{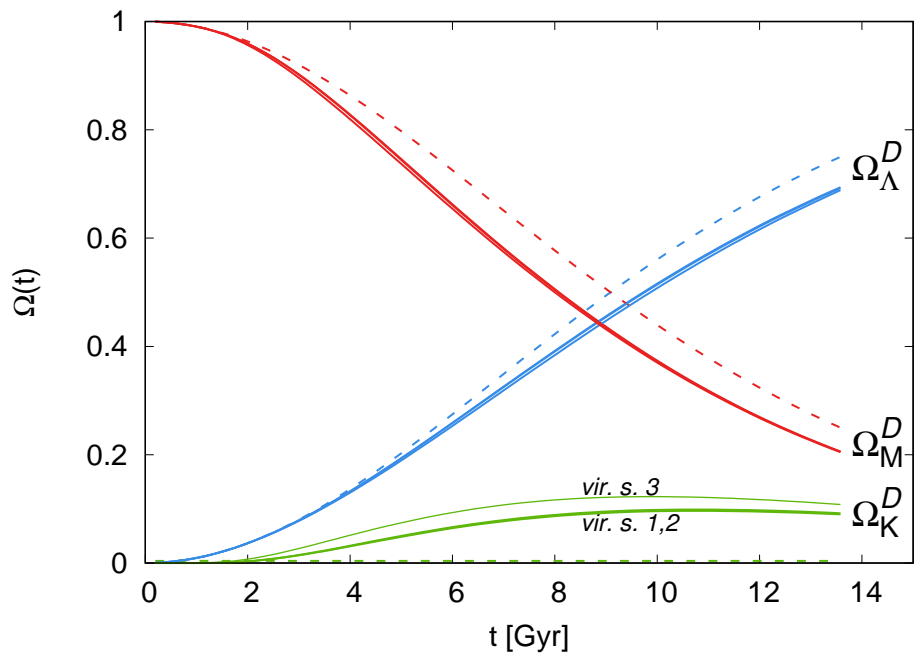}
\end{center}
\caption{ {  Emerging spatial curvature and evolution of matter and dark energy in the  Simsilun simulation (solid lines) and in the 
background  $\Lambda$CDM model of the Millennium simulation (dashed lines) }.}
\label{fig6}
\end{figure}

{  

We first start by generalising the standard cosmological parameters ($\Omega_M$, $\Omega_\Lambda$, and $\Omega_K$) to include averages and time evolution. 
Dark matter and dark energy parameters are easily generalisably and are defined as \citep{2008GReGr..40..467B}
\begin{equation}
\Omega_M^{\cal D} = \frac{8 \pi G \av{\rho}_{\cal D} } {3 H_{\cal D}^2 }  \quad {\rm ~and~} \quad
\Omega_\Lambda^{\cal D} = \frac{ \Lambda } { 3 H_{\cal D}^2 }. \label{omol}
\end{equation}
The average $\av{\rho}_{\cal D}$ is the volume average over a domain ${\cal D}$, and is defined as
\begin{equation}
\av{\rho}_{\cal D}  = \frac{ \sum_i \, \rho_i \, V_i} {\sum_i V_i}, 
\end{equation}
where $\rho_i$ is density within a given cell, and $V_i$ is a volume of a given cell, and its evolution follows from
\begin{equation}
\dot{V} = V \Theta,\label{volume}
\end{equation}
thus, apart from solving the silent universe evolution equations (\ref{rhot})--(\ref{weyt}), we also need to solve the above equation
to obtain the evolution of volume of a given domain ${\cal D}$.
Finally, the parameter $H_{\cal D} $ is the volume average Hubble parameter and is defined as 
\begin{equation}
 H_{\cal D} =  \frac{1}{3} \av{ \Theta }_{\cal D}.
\end{equation}
Relations (\ref{omol}) provide us with a  generalisation of the standard cosmological parameters $\Omega_M$ and $\Omega_\Lambda$. Firstly, this definition includes averages (as opposed to homogeneous and isotropic FLRW quantities). Secondly, these parameters are allowed to evolve with time, as opposed to the usual definition, which defines these parameters at the present-day instant.

For FLRW models, the parameters $\Omega_M$ and $\Omega_\Lambda$, constrain the spatial curvature
\begin{equation}
\Omega_K = 1 -  \Omega_M - \Omega_\Lambda.\label{triplet}
\end{equation}
The above is true even if one generalises the FLRW cosmological parameters to evolve with time, 
where $\Omega_M(t)$ and $\Omega_\Lambda(t)$ follow the definition (\ref{omol}) with
$\rho = \rho(t)$ and $H = H(t)$ being the FLRW density and expansion rate respectively.
For the inhomogeneous universe, the above formula is not exact.
To show this, we start with the Hamiltonian constraint, from which it follows that the spatial curvature is
\begin{equation}
{\cal R}= 2 \kappa   \rho + 6 \, \Sigma^2  - \frac{2}{3} \Theta^2 + 2 \Lambda, \label{hamcon}
\end{equation}
after volume averaging the above becomes
\begin{equation} \label{avcur}
\av{ {\cal R} }_{\cal D} = 2\kappa   \av{ \rho }_{\cal D} + 6 \av{ \Sigma^2}_{\cal D}  - \frac{2}{3} \av{ \Theta^2 }_{\cal D} + 2 \Lambda,
\end{equation}
and by comparing with the FLRW definition of the parameter $\Omega_K = - {\cal R}/ (6 H_0^2)$ we get  
\begin{equation}
\Omega_K^{\cal D} = -\frac{ \av{ {\cal R} }_{\cal D} } { 6 H_{\cal D}^2 }.
\label{oko}
\end{equation}
Finally, dividing eq. (\ref{avcur}) by  $6 H_{\cal D}^2$ we arrive at
\begin{equation}
\Omega_K^{\cal D} = 1 - \Omega_M^{\cal D} - \Omega_\Lambda^{\cal D} - \Omega_\mathcal{Q}^{\cal D}
\end{equation}
where
\begin{equation}
\Omega_\mathcal{Q}^{\cal D} = \frac{1} {H_{\cal D}^2 } \left( \av{ \Sigma^2}_{\cal D}  + \frac{1}{9} \av{ \Theta^2 }_{\cal D}  - H_{\cal D}^2 \right),
\end{equation}
is the kinematic backreaction. Thus, for inhomogeneous system, $\Omega_M(t) + \Omega_\Lambda (t)$ does not specify the special curvature as it does in the FLRW case. However, in most cases the kinematic backreaction
remains small  \citep{2006CQGra..23.6379B,2009PhRvD..80l3512W,2013CQGra..30q5006D,2013PhRvD..87l3503B,2013JCAP...10..043R,2017JCAP...06..025B}, and so $1 - \Omega_M^{\cal D} - \Omega_\Lambda^{\cal D}$ provides a reasonable approximation for the spatial curvature $\Omega_K^{\cal D}$.
From the Simsilun simulation, at the present-day instant, $\Omega_\mathcal{Q}^{\cal D} \approx 0.01$  for virialisation scenario 1 and 2, and $\Omega_\mathcal{Q}^{\cal D} \approx - 0.001$ for virialisation scenario 3.

The evolution of the parameters $\Omega_K^{\cal D}$, $\Omega_M^{\cal D}$, and $\Omega_\Lambda^{\cal D}$ within the Simsilun simulation
is presented in Fig. \ref{fig6} (solid lines).
The most striking results is that the spatial curvature is not zero but evolves. 
The Simsilun simulation starts with vanishing mean spatial curvature, but as the universe evolves the spatial curvature increases, and after $t \approx 10$ Gyr it peaks at $\Omega_K^{\cal D} \approx 0.1$. After this instant, the spatial curvature starts to decrease. This turnaround is related to what is usually referred to as the cosmological ``no-hair'' conjecture. The cosmological ``no-hair'' conjecture states that the universe dominated by 
 dark energy  asymptotically approaches a homogeneous and isotropic de Sitter state \citep{PhysRevD.28.2118,1991AnP...503..518P}.
Thus, when dark energy becomes dominant $\Omega_\Lambda^{\cal D} > \Omega_M^{\cal D}$
the evolution of the spatial curvature $\Omega_K^{\cal D}$ plateaus and eventually decreases.
There is a slight difference in  $\Omega_K^{\cal D}$ between the virialisation scenarios 1, 2 and the virialisation scenario 3. There reason why virialisation scenario 3 leads to a more negative spatial curvature is due to the fact that the {\em stable halo} scenario assumes vanishing shear $\Sigma = 0$.
As seen from eq. (\ref{hamcon}), $\Sigma^2$ contributes to the spatial curvature. Thus, neglecting shear within the virialisation scenario 3, means that virialised regions do not positively contribute to the spatial curvature and therefore the overall average curvature is more negative.

The emergence of the spatial curvature is consistent with findings of 
\citet{2011CQGra..28p5004R} who studied the 
global gravitational instability of the FLRW models and showed that even tiny perturbations 
in $\Omega_\mathcal{Q}^{\cal D}$ result in a non-Friedmannian evolution of the spatial curvature.
In the FLRW limit ${\cal R} \to 6k/a^2$, so $\Omega_K^{\cal D} \to  -k/(H^2 a^2)$,
where $a$ is the scale factor and $k$ is the curvature index. If 
$\Omega_\mathcal{Q}^{\cal D} \ne 0$ then the even if initially $k=0$ 
then $\Omega_K^{\cal D}$ does not stay zero (as in the FLRW models) but evolves. The results
presented in Fig. \ref{fig6} confirm these findings.

The emergence of the spatial curvature is in contrast with the evolution of the universe modelled using N-body Newtonian simulations with periodic boundary conditions imposed.
Within the comoving gauge (see below for a discussion on the gauge dependence),
the nature of Newtonian interactions combined with the periodic boundary conditions (PBC) imply \citep{1997A&A...320....1B,2017MNRAS.469..744K,2017arXiv170400703B,2017arXiv170606179R}
\[ \left.  \frac{2}{3}  \av{ \Theta^2 } - \frac{2}{3} \av{ \Theta}^2  - 6 \av{ \Sigma^2 } \right|_{Newtonian + PBC} = 0. \]
In addition, the Newtonian N-body simulations often assume that the global evolution follows the Friedmann equations, consequently both $\av{ \Theta }$ and $\av{\rho}$ follow the 
Friedmann solution, which implies via (\ref{avcur})
\[ \left. \av{ {\cal R} }\right|_{Newtonian + PBC}   = 0.\]
Thus,  within N-body Newtonian simulations the spatial curvature remains flat throughout the cosmic evolution.

The emergence of the spatial curvature is associated with the change of the evolution of other cosmological parameters as presented in Fig. \ref{fig6}.
Since $\Omega_K^{\cal D} + \Omega_M^{\cal D} +  \Omega_\Lambda^{\cal D} \approx 1$, thus the increase of the spatial curvature 
$\Omega_K^{\cal D}$ by $0.1$ results in similar decrease of the sum of $\Omega_M^{\cal D}$ and $\Omega_\Lambda^{\cal D}$. For the Simsilun simulation
this change is almost equally spread across these parameters and their amplitudes are smaller by $\approx 0.05$ compared to the $\Lambda$CDM model (dashed lines) of the  Millennium simulation (i.e. $\Omega_M = 0.25$, $\Omega_\Lambda = 0.75$ at the present-day instant).
There is a slight difference between the values of the parameters  $\Omega_M^{\cal D} $ and $\Omega_\Lambda^{\cal D}$ depending on the assumed virialisation scenario, but this difference is small and is not visible in Fig. \ref{fig6}.

The results reported in this Section and presented in Fig. \ref{fig6} were obtained within the comoving gauge. It has been a debate whether these findings are subject to a choice of a gauge.
The parameters $\Omega_M^{\cal D}$, $\Omega_\Lambda^{\cal D}$, and $\Omega_K^{\cal D}$ are evaluated via averaging over domains of constant time, and so are susceptible to a choice of slicing and therefore could be gauge dependent.
Recent results obtained by 
\citet{2015PhRvL.114e1302A,2017arXiv170609309A} 
showed that in the Poison gauge the backreaction (i.e. the change of the global evolution of the universe, for example as measured by the parameters $\Omega_M^{\cal D}$, $\Omega_\Lambda^{\cal D}$, and $\Omega_K^{\cal D}$)
is negligible small, while in the comoving gauge (as in this paper) the backreaction can be large.

While it is not surprising that one can  choose such a slicing where the backreaction vanishes, the question remains, which gauge is the most appropriate for studying the properties of the universe.
Ultimately, the solution to this problem will be provided by ray tracing and evaluating cosmological observables in a gauge invariant way.
However, it is interesting to notice that recent measurements of matter density 
based on weak leaning (DES) in the low-redshift universe show
 $\Omega_M = 0.264^{+0.032}_{-0.019}$  \citep{2017arXiv170801530D} which is lower than the CMB preferred value of $\Omega_M = 0.308 \pm 0.012$ \citep{2016A&A...594A..13P}.
Similarly, supernova observations (JLA) alone suggest lower values: 
 $\Omega_M \approx 0.22$ and  $\Omega_\Lambda \approx 0.55$ with very large uncertainties.
It is only when these measurements are combined with high precision (i.e. small uncertainty) CMB data 
that they shift toward higher values: 
the DES data when combined with the Planck, SDSS, 6dF, BOSS, and JLA constraints push the matter density from  $\Omega_M = 0.264^{+0.032}_{-0.019}$ 
to  $\Omega_M = 0.301^{+0.006}_{-0.008}$; and the JLA supernova data when combined with CMB and BAO shift the matter density and cosmological constant to $\Omega_M = 0.305 \pm 0.01$ and 
$\Omega_\Lambda = 0.693 \pm 0.01$.

While measurements of matter density $\Omega_M$ at low redshift are subject to number of systematics, and may not directly correspond to $\Omega_M^{\cal D}$, the results of the Simsilun simulation suggest that there should be a discrepancy between high and low redshift measurements.
To theoretically estimate this discrepancy to high confidence one needs to develop ray tracing codes and evaluate observables in a gauge independent way.
To empirically measure these   discrepancies one requires high precision and high accuracy measurements free of any major systematics both at low and high redshifts.

}

\section{Conclusions}\label{conclusions}

This paper presents the results obtained from the code \textsl{SIMplified SILent UNiverse (simsilun)}.
The code solves the Einstein equations under the approximation of the `silent universe' and evaluates the evolution of a cosmological system based on 4 scalars: density $\rho$, expansion $\Theta$, shear $\Sigma$, and Weyl curvature ${\cal W}$. 
In addition the code \textsl{simsilun} simplifies the procedure for setting up the initial conditions (hence the name `Simplified Silent Universe'),
and allows to set up the model using only the initial density contrast. 
In this paper the initial conditions were set using the Millennium simulation, which provided a coherent set of initial conditions together with realistically appearing cosmic structures (cf. Fig. \ref{fig4}).

{ 
The simulation obtained this way, referred to as the Simsilun simulation, was employed to study the evolution of cosmological properties $\Omega_M^{\cal D}$,  $\Omega_\Lambda^{\cal D}$ (cf. eqs. (\ref{omol})) and $\Omega_K^{\cal D}$ (cf. eq. (\ref{oko})).
Compared to the $\Lambda$CDM model (i.e. the background model of the Millennium simulation) the spatial curvature is not flat but evolves  from 
spatial flatness of the early universe to $\Omega_K^{\cal D} \approx 0.1$ at the present-day instant (see Fig. \ref{fig6}).
The emergence of the spatial curvature is associated with both $\Omega_M^{\cal D}$ and
$\Omega_\Lambda^{\cal D}$ being smaller by approximately $0.05$ compared to the $\Lambda$CDM model (note that in the $\Lambda$CDM model we have $\Omega_M + \Omega_\Lambda = 1$, whereas for the Simsilun simulation $\Omega_M^{\cal D} + \Omega_\Lambda^{\cal D}  \approx 1 + \Omega_K^{\cal D}$).

The code \textsl{simsilun} is publicly available via the \textsl{Bitbucket repository}\footnote{\url{https://bitbucket.org/bolejko/simsilun}} and is distributed under the terms of the GNU General Public License. The code can easily be modified to include further effects, such as light propagation \citep{2010JCAP...03..018R,2011MNRAS.412.1937B,2012MNRAS.426.1121C,2012JCAP...05..003B,2014JCAP...03..040T,2014PhRvD..90l3536P}, modelling of inhomogeneous non-symmetrical cosmic structures \citep{2006PhRvD..73l3508B,2007PhRvD..75d3508B,2012PhRvD..85h3502I,2012PhRvD..86l3508P,2015PhRvD..92h3533S,2016JCAP...03..012S}, and the impact of the local cosmological environment on astronomical observations \citep{2016JCAP...06..035B}.

The evolution of the silent universe upon which the code is based is subject to several approximations and limitations. The silent approximation assumes vanishing pressure gradients, vanishing rotation, and vanishing magnetic part of the Weyl tensor. These assumptions are valid and preserved by linear and second order perturbations, but are expected to break down during a highly nonlinear stage of a collapse  \citep{2002PhRvD..66l4015E}. 
 At this stage of the evolution also
shell crossings start to develop and the fluid approximation is expected to break down.
The code \textsl{simsilun} deals with this problem by forcing collapsing domains to virialise. Three different virialisation scenarios has been implemented and the results seems to be very weakly dependent on the choice of the virialisation mechanism.
Finally, the results were obtained within the comoving gauge and therefore are subject to the limitations and applicability of the comoving coordinates \citep{2015PhRvL.114e1302A,2017arXiv170609309A}.

It seems that these limitations can only be properly addressed with full numerical relativistic cosmology. While the community is working towards such solutions, they are still not available yet \citep{2017IJMPD..2630011B}
Comparison between different existing codes, such as the \textsl{simsilun} (this paper), \textsl{gevolution} \citep{2016JCAP...07..053A}, and codes based on the BSSN formalism \citep{Bentivegna:2015flc,Mertens:2015ttp,2017PhRvD..95f4028M}
could provide insight into properties of relativistic systems such as our universe.
}

%\section*{Acknowledgements}
\ack

This work was supported by the Australian Research Council through the Future Fellowship FT140101270. 
The Millennium Simulation databases used in this paper and the web application providing online access to them were constructed as part of the activities of the German Astrophysical Virtual Observatory (GAVO). Computational resources used in this work were provided by the ARC (via FT140101270) and the University of Sydney HPC service (Artemis). 
 Finally, discussions with Boudewijn Roukema, Daniel Price, Paul Lasky, Hayley Macpherson,
  and participants of the \textsl{Inhomogeneous Cosmologies} (\url{http://cosmo.torun.pl/CosmoTorun17}) are gratefully acknowledged.

\bibliographystyle{mnras_openaccess}
\bibliography{simsilun_rev}

\end{document}